\documentclass[showpacs]{revtex4}
\usepackage{graphicx}
\usepackage{latexsym}
\usepackage{amsmath}
\usepackage{amssymb}
\usepackage{amsfonts}
\usepackage{amsthm}
\usepackage{epsf}
\usepackage{graphicx}        
\input epsf
\def\ll{\label}
\def\re{\ref}
\def\c{\cite}

\def\r1{(\ref{$1})}

\def\ep{\epsilon}

\def\ba{\begin{array}{c}}

\def\ea{\end{array}}

\def\bet{\beta}

\def\l{\left}
\def\l({\left(}
\def\r){\right)}
\def\r{\right}

 \def\be{\begin{equation}}
\def\bc{\begin{center}}
\def\ec{\end{center}}
\def\bit{\begin{itemize}}
\def\eit{\end{itemize}}
\def\ee{\end{equation}}
\def\ed{\end{document}}
\def\bea{\begin{eqnarray}}
\def\eea{\end{eqnarray}}
\def\efr{\end{flushright}}

\begin{document}
\title{ { Asymmetric  Skyrmion lattice   
  in helimagnets 
 }}
\author{ Anjan Kundu\\
 Theory Division,
Saha Institute of Nuclear Physics\\
 Calcutta, INDIA\\
{anjan.kundu@saha.ac.in}}
\begin{abstract}
  Intricate spin textures     in  helimagnets,  
    identified as stable topological Skyrmions,  were observed
 experimentally, where Skyrme lattice was supposed to exhibit symmetric
 structures in the ground state.
  We show the possibility  of asymmetric Skyrmions in a helimagnetic model, for
 individual Skyrmion as well as for the hexagonal Skyrme crystal with higher
charge, as perturbative deformation and stabilization of exact ferromagnetic
solitons.  Such   nonsymmetric  configurations for the  Skyrme lattice,
predicted here theoretically, need to be verified in precision experiments.

\end{abstract}

 \maketitle


\section{ Introduction}

 Unusual magnetic behavior of  helimagnet MnSi
\c{unusual}, which triggered the suspicion, that the magnetic states in such
helimagnetic materials might be of topological origin
\c{topological,top2,bogdanNature06,bogdanMnsi}, is confirmed as Skyrmions in quantum
Hall (QH)
 \c{QHexperim} and small angle  neutron    scattering (SANS) experiments \cite{nutronScattExp}.
  These experiments, though made pioneering observations,
  could give only  indirect evidence for the existence of Skyrmion spin
textures, due to their confinement in
 the momentum space.  However, more recently,  in a real space
experiment with  Lorentz transmission electron microscopy (TEM), direct  photographic evidence
of Skyrmion crystals was
obtained in a thin film of helimagnet Fe$_{0.5}$Co$_{0.5}$Si,   on a plane perpendicular to
the applied magnetic field
 \cite{LorentzExp}. Skyrmion crystals were found to be  in a  beautiful
  hexagonal form, with the localized individual
Skyrmions serving as its constituent molecules, extended over  lattice spacing
of 30  nm range.

For theoretical description of these fascinating experimental  observations in helimagnets, 
 the accepted basic model is given by a ferromagnetic spin exchange
Hamiltonian in
 combination with a Dzyaloshinskii-Moriya (DM) interaction
\c{bogdanMnsi,LorentzExp,heliModel,heliModelPRL10}.  It is argued,
that the occurrence of topological solitons in helimagnets is due to the
competing forces between the ferromagnetic  and the effective DM
interactions, where  the ferromagnetic spin exchange
 tends to align the neighboring spins parallel to each other, while
  the DM spin-orbital interaction  with its broken inversion symmetry, 
orients them to be  mutually perpendicular, resulting  to a 
 helical order of topological origin .

Since the experimentally  observed Skyrmions are found to have extended
nature,
 slowly varying in comparison with the 
lattice structure of the original magnetic crystals, 
   the long wavelength limit is justified and 
at low energy  and weak DM coupling we can consider 
 the continuum approximation. At this  semiclassical limit an
intriguing topological property sets in, with a spatially dependent magnetization
 as  a unit vector, wrapping  around a 2-sphere, while  the 2d coordinate  space,
 due to the fixed orientation of the spins at space-infinities, 
  compactifies  to another 2-sphere, 
 inducing a 
sphere to sphere  mapping.
 The degree
of this mapping,   counting the number of  times 
the Skyrmion 
 magnetic field sweeps the target sphere, when the coordinate space is
 covered once, defines the integer valued topological charge $Q=N$
\cite{Skyrme}.
 
It is a remarkable fact, that at the continuum limit, the ferromagnetic
Hamiltonian itself, defined on a 2d plane,
  allows  exact Skyrmion solutions with arbitrary integer
charge $N $, which can be  
 linked to the holomorphic  functions \cite {BP}.  However, such solutions
 exhibiting in general a  noncircular symmetry,  posses a scale invariance
property, which does not allow to fix the Skyrmion size and makes the
Skyrmions in a ferromagnetic model unphysical metastable states. 
However, the addition of a DM type interaction with broken symmetry can
change this picture significantly by providing the necessary scale through
its coupling parameter and stabilizes  the magnetic Skyrmions, that have been
observed in several recent experiments in different helimagnetic materials. 
However,
 at the theoretical level, the addition of the DM Hamiltonian 
 can not sustain the analytic solutions obtained in the original
ferromagnetic model and at the same time introduces
  a high level of  asymmetry to the resulting equations, due to
 explicitly broken inversion symmetry in the  DM interaction.  Therefore,
the appearance of Skyrmion spin patterns with noncircular symmetry seems to
be more generic in such systems, in particular for describing the Skyrme
crystals.  
However based on the  experimental data, available so far,
 in  observing the magnetic Skyrmions in
helimagnets,  any definitive statement about the 
symmetry and isotropy of the individual Skyrmions located in the Skyrme
lattice seems to be difficult to make. In  QH  experiments the indirect evidence of
Skyrmions was obtained in the form of an unusual magnetic field giving no
details about the structure of the soliton. In
 SANS experiment
, though the situation was improved with the detection of the hexagonal form of
Skyrme crystals through a 6-fold symmetry, any symmetry statement about the
individual Skyrme molecules was beyond the scope of this experiment.
Finally in
 recent thin film Lorentz TEM  experiment   
    images of extended Skyrme molecules inside  hexagonal Skyrme crystals 
were obtained in  real space, revealing  
 much detailed structures of these objects appearing in a helimagnetic
material. However, even from this best
available data, one can not  confirm with certainty 
about the nature of   symmetry and isotropy of the individual Skyrmions,
since the color-code of the images shows directional
asymmetry and  color inhomogeneity (see Fig. 1e-f of \cite{LorentzExp}).
 Therefore, for a clearer picture about the  precise symmetry of the
Skyrmions and the exact directional orientations of the magnetic field lines
one has to look for further  finer and precision experiments.

Nevertheless, at the theoretical level, most of the models tend to support
the symmetric form of the individual Skyrmions and their isotropy inside a
Skyrme crystals, neglecting the inter-Skyrmion interactions.  For example,
in a numerical simulation by minimizing the discretized model Hamiltonian
using the Monte Carlo method,
 circular symmetric Skyrme molecules  forming a  Skyrme crystal in symmetric
form   have been obtained \cite {LorentzExp}.
Similarly, in the prevalent theoretical models  \c{bogdanMnsi,heliModelPRL10}
 a circular symmetric ansatz is assumed for the solution of individual
Skyrmions. However, since  the general nature of the governing 
 Euler equations are highly asymmetric, such a
symmetric assumption put strong restrictions on    the
corresponding  solutions,  confining it    
to  the unit charge sector and imposing  a fixed initial
phase shift,
 which limits    the spin pattern only to a    particular
orientation ($D_n $ type) \c{bogdanMnsi,heliModelPRL10}.
 
It is important to stress, that since the individual Skyrmions as
well as the Skyrme crystals  in general should be  solutions of the energy
minimizing equations, they  should  ideally be  derived from the Euler
equations  dictated by the governing helimagnetic model without restriction
of symmetry, orientation and charge.
  However, no satisfactory theoretical proposal seems to have been offered
to this challenging problem of describing the hexagonal Skyrme crystal
structure with noncircular symmetry, as a solution of the underlying
equations with higher topological charges, which would induce nonlinear
interactions between the constituent Skyrme molecules.  It is indeed a
difficult problem to solve the governing equations, representing highly
nonlinear coupled partial differential equations allowing no separation of
variables.

 Current studies on the helimagnetic model avoid
this difficult problem by completely neglecting the inter-molecular  interactions
between the Skyrmions,  by assuming a circular cell approximation
\c{bogdanMnsi,heliModelPRL10}, though  
such a linear  superposition of individual
 solutions  does not represent  a solution
 of the governing  nonlinear equations. Moreover,    
such a configuration  as a equidistant collection of individual Skyrmions
  can not be reproduced    as a limiting case of a  general solution
with a higher charge.
 
Other investigations on the Skyrme lattice in the  helimagnetic model focus
usually  on different other aspects like the
  effect of current and temperature gradients on the rotating of Skyrme
lattice \cite{SKLrot},
pinning of  Skyrmion through inhomogeneity of magnetic exchange
 \cite{SKpin}, Skyrmion spin lattice  coupling with electric dipole
\cite{SKLedipol} etc., leaving aside  the basic  problem of the
formation of Skyrme crystals as a topological solution.

Our aim here is to focus on the  above posed   problem of constructing
a Skyrme crystal as a solution of the helimagnetic equations. However, we  
bypass the  direct problem, which is hard to solve
 and obtain   the solutions with higher topological charges linked to the
Skyrme crystal, by looking first for the Skyrmions appearing in the pure
ferromagnetic model (FM)
 and subject  them subsequently, to the DM interaction perturbatively, by
 assuming the coupling to be  weak.   This 
stabilizes  the solitons  through DM coupling and  goes beyond the
limitations of the  earlier studies  by allowing  Skyrmion solutions with higher charges
and  more general  initial  phase admitting 
 different types of  orientations for the spin texture.
   This would mean, however
that one has to deal
 with noncircular symmetric  fields without separation of   variables, which 
 unlike earlier proposals, is a more difficult task  to handle. 
Nevertheless, this approach   solves to some extend the 
  problem of formation of  hexagonal  Skyrme crystals within  the helimagnetic model
 with weak DM coupling, 
 by finding a perturbed Skyrmion solution with charge $Q=-7 $, and 
without the usual circular cell approximation.

.
\\
\section { The Model}
The basic Helimagnetic model, as motivated  above, is given by the
Hamiltonian  \bea    
& & H= J \ H_{FM}+D \ H_{DM}, 
  \ll{Hheli} \\
  \mbox {with} \ & & \ H_{FM}= \frac 1 2 \int d^2x (\nabla {\bf M})^2, \ll{FM}  \\  
& & H_{DM} = \int d^2x({\bf M } \cdot [\nabla \times {\bf M}])
 ,\ll{DM} \eea
where   ${\bf M } $  with ${\bf M}^2=1 $ is the magnetization vector  and  $H_{FM} $,
  $H_{DM}$ are  the competing  ferromagnetic  and  DM interactions,
  respectively \c{bogdanMnsi,LorentzExp,BakJensen}.
Since the magnetization, taking values on a  2-sphere is restricted to a 2d
plane, the system may be described   
by  two spherical  angles  $\beta (\rho,\alpha) $ and $\gamma (\rho,\alpha) $   
defined through  polar coordinates $ (\rho,\alpha)  $ with the magnetization 
 components given as  $M^1\pm i M^2= \sin \beta e^{ \pm i \gamma}, \
M^3=\cos \bet . $
 The  helimagnetic Hamiltonian  (\ref{Hheli})
 may  be   expressed  through these  angle variables for convenience as
\bea 
& & H_{FM}=\frac 1 2 \int d^2x((\nabla \beta  )^2+\sin ^2 \beta (\nabla \gamma )^2),
\ll{CFM} \\  
 & &  H_{DM}= \int d^2x (  \sin (\gamma-\alpha)(\partial_\rho\beta-\frac 1 {2 \rho}  \sin
2 \beta \ \partial_\alpha \gamma)  -  \cos (\gamma-\alpha)(\frac { \partial_\alpha\beta} \rho+\frac 1 {2 }
  \sin 2 \beta \ \partial_\rho \gamma))
. \ll{CDM}\eea

The associated topological charge  takes the
explicit form
\bea
Q= \frac {1}{   4  \pi } \int d^2 x \sin \beta 
 |[{\nabla \beta } \times {\nabla \gamma } ]| 
, \ll{Qba}\eea


\subsection{The Euler equations}

Note that the energy minimizing   Euler equations
  for the fields $\beta $ and $
\gamma $, may be  derived  in the static case  from the model
Hamiltonian (\ref {CFM}, \ref{CDM})  as 

\bea
 && \nabla^2\beta - \frac 1 2  \sin 2 \beta (\nabla \gamma)^2
+ 2 \epsilon M_\beta (\beta,\gamma) =0 , \ \  \ \   \epsilon= \frac D J  \ll{betaEq} 
 \eea for angle $\beta(\rho,\alpha) $ and   
\bea
 && \sin 2 \beta (\nabla\beta \cdot \nabla \gamma) +
\sin^ 2 \beta (\nabla^2 \gamma)+2 \epsilon M_\gamma
(\beta,\gamma)=0,  \ll{gammaEq} 
\eea
and with respect to angle $\gamma(\rho,\alpha) $ , where   
the additional DM terms are

\bea
&& M_\beta (\beta,\gamma) \equiv
 \ \sin^ 2 \beta( \partial_\rho \gamma \cos (\gamma-\alpha) + \frac {1} {\rho }
\partial_\alpha \gamma \sin (\gamma-\alpha))   \ll{EqDMbeta}
\eea and \bea
&& M_\gamma (\beta,\gamma) \equiv 
- \ \sin^ 2 \beta( \partial_\rho \beta \cos (\gamma-\alpha)+  \frac {1} {\rho } 
\partial_\alpha\beta \sin (\gamma-\alpha)
=0. \ll{EqDMgamma}
\eea
Note that these   equations for finding the field
configurations  minimizing the Hamiltonian, represent 
 highly nonlinear
  inhomogeneous  coupled partial
differential  equations in  two variables and  are difficult to solve in
general.
A simple possibility to overcome this difficulty, as adopted in most of the
earlier studies \cite{bogdanNature06,bogdanMnsi,heliModelPRL10}, is to focus on
an individual Skyrmion
 with a circular symmetric   ansatz
  \be \beta=\beta(\rho), \  \gamma =N \alpha +\alpha_0,
 , \ll{rhoalpha0} \ee with    initial
phase $\alpha_0 $, as an arbitrary   constant and $N$ as an arbitrary integer. 
 Specific values of $\alpha_0 , $
represent  precise crystallographic forms, describing the spin  orientation in the
related magnetic pattern  \c{bogdanMnsi}.  A closer look however reveals that
for the circular symmetric ansatz (\re{rhoalpha0}) to work,
 the equation (\re {gammaEq}) for the angle
$\gamma $  must be zero, which  due to
$\partial_\alpha \beta=0 ,\ \partial_\rho \gamma =0$ 
 reduces to the vanishing of 
$M_\gamma$ 
 (\ref {EqDMgamma}) or in turn  to the vanishing condition for the
term 
$$ \ \sin^ 2 \beta \  \beta^{'} \cos ((N-1)\alpha+ \alpha_0)
\ . $$ 
Clearly  it  becomes zero (for $\beta^{'} \neq 0 $), when the cosine function vanishes
  under the combined condition: $N=1 $ and
 $\alpha_0=\pm \frac  \pi 2 $. Therefore, for the validity of the  symmetric ansatz
(\re{rhoalpha0}) for the  Skyrmions in helimagnets, one
gets a   solution   restricted only to    unit
topological charge $Q\equiv N=1 $ and with   fixed value for  the initial phase
linked  to  $D_n $ type of magnetic  pattern \cite{bogdanNature06,bogdanMnsi,heliModelPRL10}.
Therefore,    though this is a convenient way 
to achieve  separation of variables    in  
 a  coupled nonlinear equation, it can not get   generic Skyrmion solutions
 with higher topological charges
$Q=N >1 $
and misses the possibility of finding other crystallographic forms for other
values of
 $\alpha_0 $, involving noncircular symmetric solutions.
 More importantly, 
 such a symmetric  ansatz can not   describe the  Skyrmion lattice structure
 having multiple centers and higher
topological charge,  including the  Skyrme
crystals  in hexagonal form, as a solution of the full set of 
 governing equations (\ref{betaEq},\ref {gammaEq}).  Nevertheless, a formal
insertion of such circular symmetric Skyrmions at each lattice site,
sometimes  with 
    anisotropic terms
like $h \ M^3 , K \ (M^3)^2$ etc.
added to the helimagnetic
model (\ref{Hheli}),  was proposed theoretically  for
describing  a Skyrme crystal  \cite{bogdanMnsi}, which however  
 neglects  completely the  interaction between 
the  Skyrme molecules,  under 
 circular cell approximation.
\section { The Approach}
Going
 beyond the conventional  circular cell
approximation, we    intend to  look for more general noncircular 
symmetric solutions  by considering  interactions between the individual Skyrme
molecules, 
consistent with the   coupled  Euler equations
(\re{betaEq}-\re{gammaEq}), minimizing the  helimagnetic model. 
However, since such  asymmetric   solutions are difficult to obtain in
general,  we adopt a bypass route by considering the
deformation of   Skyrmion 
 solutions of   pure  ferromagnetic  model  (\ref{FM}),  by  switching  on  
  the DM interaction (\ref{DM}) perturbatively, by taking the
   DM coupling $D$ weaker in   comparison with    the ferromagnetic exchange
coupling $J$ with
 $\epsilon \equiv D/J  $ as a   small parameter.
  The   assumption of a
weak relative coupling $ D/J, $ however  has gone already in
the justification of
 the    continuum approximation  for the  present
model (see  e.g. \c{LorentzExp}).  Note, that this approach also brings in
the necessary scaling parameter for stabilizing the Skyrmions.

\subsection{Skyrmions in ferromagnetic model}
 In the first step,  we focus  on the nonlinear Euler equations associated
with the ferromagnetic model alone, without the DM interaction and look for
the exact Skyrmions with higher topological charge, which in general would have
noncircular symmetry.  In the next step, this exact solution is
subjected to the DM interaction perturbatively, where though the perturbing
fields do not allow separation of variables, the governing equations 
become linear, allowing numerical solutions. 
Therefore, we start with suitable solutions $\beta_0 (\rho,\alpha), \gamma_0
(\rho,\alpha)
 $ for the ferromagnetic model (\ref{FM}) (or (\ref{CFM})), satisfying the
related Euler equations (\re{betaEq}-\re{gammaEq}), at $D=0 $.
 It is remarkable, that the topological Skyrmions  for this model
 can be given by  the celebrated exact  solutions found way back by   Belavin
and Polyakov (BP) \c{BP}.
A deep theoretical concept and  beautiful geometrical  picture go into this
construction, where an  important lower bound for the energy
(\ref{CFM}) is revealed  through the   topological charge $Q$
(\ref{Qba}):
$H_{FM}
\geq   4  \pi |Q| $. Note that due to the conservation of  charge 
the lower bound guarantees, that the finite energy topological solitons can
not decay into the vacuum or any other topological  states. Interestingly, 
the lower bound  is saturated under the so called {\it self duality} 
  condition
\be \partial_\rho \beta_0=\pm  \frac 2 \rho \  \sin \beta_0  \partial_\alpha
\gamma_0
, \  \partial_\alpha \beta_0=\mp 2  \rho \  \sin \beta_0  \partial_\rho
\gamma_0
  \ll{Bog} \ee
 and since  this  is the minimizing
condition for the     energy  functional, the corresponding   field
configuration  also   becomes a solution of the  associated Euler
equations ((\re{betaEq}-\re{gammaEq}) at $D=0 $), which are 
 our current focus. Note, interestingly that these Euler equations 
are actually  solved
without  solving them directly,
 but by solving   the self-duality condition and therefore, 
for finding the Skyrmion solutions in  pure ferromagnetic
 model, one has to consider  only the solution of   (\re{Bog}), which
remarkably can be  solved  exactly. Deep reason behind this intriguing fact is
that the    relations   (\re{Bog}) for  attaining the energy minimum represent the
 well known Cauchy-Riemann (CR) condition (expressed in the  polar
coordinates), required  for the  analyticity of the  complex  function $f(z),  $
linked   through the stereographic projection
\be 
 f(z)= \frac {M^1+iM^2} {1+M^3} \equiv \tan {\frac {\beta_0} 2}e^{i\gamma_0}
.\ll{stereo} \ee
 Consequently, as shown in \cite{BP},  any analytic function of the form 
\be  f(z)= \frac {\prod_i(z-z_i)^{n_i} } {\prod_j(z-z_j)^{n_j}}, \ N=\sum_i
n_i-\sum_jn_j
 , \ll{BPsol} \ee
 would be an exact Skyrmion solution of the ferromagnetic model with topological charge $
Q= N$, allowing a  scale invariance: $z \to \lambda z  $.  

Our strategy in constructing the Skyrmions for the Helimagnetic model,  is
to take    the
 ferromagnetic soliton solutions  (\re{BPsol}) as the unperturbed Skyrmions $\beta_0,
\gamma_0 $ through mapping (\ref{stereo}), giving
the angle variables as inverse functions
\be \beta = 2 \tan ^{-1} (|f |), \ \gamma= {\rm arg} f. \ll{betagamma} \ee
 These are guaranteed to   satisfy the self-duality (CR condition) (\re{Bog}) and
consequently,  
 the Euler-equations (\re{betaEq}-\ref{gammaEq}) for $D=0 $, as an   exact
solution.
 Therefore,  for describing an individual Skyrmion we may consider the
simplest case with $Q=-1$,
  obtained from  (\re{BPsol}) through a reduction $z_j=n_i=0, \ n_j =1 $ 
giving an exact    solution $f(z)\equiv f_1(z_1=0)= \frac 1 z \equiv \frac 1 \rho e^{-i \alpha }
$, where scale transformations $\rho \to \rho_0 \rho , \alpha \to \alpha +\alpha_0$ are
allowed. This  isolated Skyrmion  of the ferromagnetic model, 
linked to  the magnetization component  \be 
 M^3=\cos \beta= \frac {1- | f_1|^2} 
{1+| f_1|^2}= \frac {\rho^2- 1} 
{\rho^2+ 1}, \ll{f1} \ee having  a perfect circular symmetry is   shown  in Fig.  1a
 in the graphical form.
 
For constructing    Skyrmion lattice
 in a ferromagnetic model, a naive  way would be  to insert  the same
circular symmetric BP 
Skyrmion  with $Q=-1$, constructed above,   at each Skyrme lattice site 
on the 2d plane,  separated uniformly from
its neighbors  at a distance of $a_s $. This would 
give the   magnetization in the form
$ M^3=\cos \beta_0= \sum_j \frac {1-| f_1(z_j)|^2} {1+|
f_1(z_j)|^2}, $ neglecting  the nonlinear  interactions between the Skyrme
molecules, induced by the governing Euler equations. 
  Note, that a  similar construction  was adopted
 for the   Skyrme crystals, but for the 
helimagnetic models,  in most of the current  studies assuming a circular
cell approximation.
However, unfortunately such a formal construction of Skyrme lattice using 
superposition of individual Skyrme solitons is {\it not} a solution of the
 energy minimizing equations dictated by the model Hamiltonian, both for
the pure ferromagnetic as well as for the helimagnetic models, since due to  
nonlinearity of the governing equations  the superposition principle of individual
solutions does not hold. Therefore any  solution for the  Skyrme crystal   with higher charges
must satisfy the
energy minimizing Euler equations, where the
nonlinear interactions between the individual Skyrmions  with unit
charge, sitting at each lattice site, can not be neglected.    

For constructing such a Skyrme crystal solution,
 we have to look  therefore for Skyrmions  with a higher topological charge,
which for the pure ferromagnetic model can be reduced   
 again from the exact BP solution, which  naturally
can  no longer exhibit circular symmetry.  For the construction of a
possible solution for the 
 Skyrmion crystal  in   hexagonal form, within the  ferromagnetic model,
 we may consider an exact    soliton 
with  topological charge $Q=-7 $, obtained  from the BP solution (\ref{BPsol})
under certain  reduction 
of  parameters $z_j, n_j, j=[1,7] $ etc.,  chosen  suitably at seven  
centers of the  equidistant lattice sites, to form a hexagonal lattice with
equal sides $a_s. $ In explicit form this solution may be given simply by  
 \be f_7^{-1 }(z)= \prod_{j=1}^7  {(z-z_j)}= - a_s^6 \rho  e^{i \alpha }+\rho^7  e^{i 7 \alpha }
.\ll{f7} \ee
This exact   solution
  may be  linked  to the hexagonal  Skyrme crystal
 pattern for the angle
variables $ \beta_0, \gamma_0 $  through the magnetization
 component \be M^3=\cos \beta_0= \frac {1-| f_7|^2} {1+|
f_7|^2}, \ \mbox{and} \    \gamma_0= -{\rm arg } f_7(z), \ll{Syrm7} \ee 
 which is obviously noncircular symmetric and expressed through
 BP solution (\ref{f7}), as  graphically
represented in 
Fig. 1b.
This topological soliton solution with higher charge describing the 
hexagonal Skyrme  crystal in a ferromagnetic model 
is   guaranteed to satisfy the required  
energy  minimizing equations,   where  parameter   $a_s $ may  serve  as the Skyrme
 lattice constant fixing the size of the
crystal. However  this parameter  remains  arbitrary due to the scale invariance
 $\rho \to \rho_0 \rho $ of the solution,  which also makes  the Skyrmion 
to vanish    at $\rho_0 \to 0, $ 
  inducing   the well known unphysical
 metastable  Skyrmionic states  in a pure   ferromagnetic model.
However, the additional DM interaction (\ref{DM}), which is included in the
helimagnetic model could stabilize the Skyrmion solutions by introducing the
required scaling  in the system through coupling parameter $D$, as we will see below.   
  Note, that
though in Fig 1a an isolated Skyrmion with unit charge exhibits perfect
isotropy and circular symmetry, such individual symmetries no longer remain
intact in the Skyrme solution of crystalline form with higher charge $Q=-7 $, as
shown in Fig 1b, due to interactions between the Skyrme molecules induced
unavoidably by the underlying nonlinear equations. 
 This characteristic
behavior of the Skyrme crystals with higher charges exhibiting anisotropy
and noncircular symmetry, as obtained here  in the
ferromagnetic model, is  expected to  remain prevalent  also in the helimagnetic model
with the inclusion of an additional symmetry breaking DM term (\ref{DM}), at
least in the excited states.
 This is the  emphasis of our proposal in what  follows.   
\begin{figure}[h!]
 \includegraphics
[width=6cm,height=4 cm]
{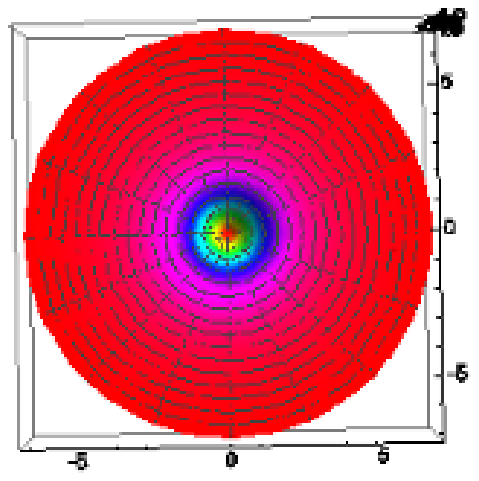}
\quad \quad \quad
 \includegraphics
[width=6.cm,height=4. cm]
{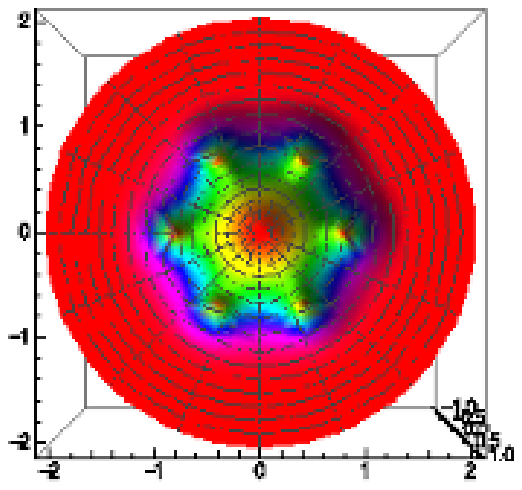}
\caption{ Magnetization component $M^3=\cos \beta_0 $ in pure ferromagnetic  model
 (\ref{FM}) for a) individual Skyrmion with $Q=-1 $ and b) hexagonal  Skyrmion
crystal with $Q=-7 $. 
   }
\end{figure}

\subsection{Skyrmions in helimagnetic model}
 Our next step  is to consider the  helimagnetic model (\ref{Hheli}) by
switching on  the DM interaction  (\re{CDM})
 to the ferromagnetic model (\re{CFM}) investigated above. 
For finding the spin texture in this helimagnetic model described by 
the    Skyrmion solution, one has to find the energy minimizing configuration
given by the  
 the solutions of the  associated  Euler equations for the fields $\beta $ and $
\gamma $ derived from (\re{CFM}-\re{CDM}).
However, as discussed above, since these coupled highly nonlinear partial differential equations
 (\re{betaEq},\re{gammaEq}) 
 for the  two fields  involving both the polar coordinates without separation of
variables, 
are difficult to solve in general for $D \neq 0 $, our strategy would be to    treat the system
perturbatively  by  considering  $ \ep=D/J, $ the relative coupling between
the DM and the ferromagnetic exchange interaction, as
 a  small parameter.
  The
perturbative solution  therefore  
may be given by
 \be \beta (\rho,\alpha)=\beta_0 (\rho,\alpha)+  \ep \
\beta_1 (\rho,\alpha), \ \gamma (\rho,\alpha)=\gamma_0 (\rho,\alpha)+  \ep \
\gamma_1 (\rho,\alpha), \ll{Pbetgam} \ee 
in the first order of approximation,  where $\beta_0 (\rho,\alpha), \gamma_0 (\rho,\alpha)
 $ are the  unperturbed solutions  induced by  the ferromagnetic  Hamiltonian
(\ref{CFM}), while $\beta_1 (\rho,\alpha), \gamma_1 (\rho,\alpha)
 $ are the  deformations suffered, when the DM interaction  (\re{CDM}) is switched on,
perturbatively. The parameter $D $  also serves as the scaling parameter, breaking the scale
invariance of the unperturbed 
Skyrmions and thus providing the required stability to the soliton solutions.
 Recall, that we have derived already the set of solutions
$\beta_0 , \gamma_0$ in the analytic form for  both  the individual Skyrmion
and the Skyrme crystal, as shown in Fig 1. Therefore, it remains now to   
 extract the deforming
solutions $\beta_1 , \gamma_1$. For this  
  we insert the perturbative ansatz (\ref{Pbetgam}) in the 
full set of helimagnetic  equations (\re{betaEq},\re{gammaEq})   and derive
the corresponding linear set of equations 
 in the first
order of approximation  $O(\epsilon) $:
\bea &&  \nabla^2\beta_1 - \sin 2 \beta_0 (\nabla \gamma_0 \cdot \nabla
\gamma_1)-\beta_1 \ \cos 2 \beta_0 (\nabla \gamma_0)^2
+ \frac 1 2 M_\beta (\beta_0,\gamma_0)=0  , \ll{PertbetaEq}  \\
 && \sin 2 \beta_0 (\nabla\beta_0 \cdot \nabla \gamma_1+\nabla\beta_1 \cdot \nabla
\gamma_0  + \beta_1 \nabla^2 \gamma_0) +\beta_1 
\ \cos 2 \beta_0 (\nabla\beta_0 \cdot \nabla \gamma_0)+
\sin^ 2 \beta_0 (\nabla^2 \gamma_1)+ \frac 1 2 \ M_\gamma (\beta_0,\gamma_0)=0,  \ll{PertgammaEq}
\eea
 where $ M_\beta (\beta_0,\gamma_0),
 M_\gamma (\beta_0,\gamma_0) $ are as in (\re{EqDMbeta}, \re{EqDMgamma}).
 However, due to the
anisotropic and symmetry breaking contribution of the DM interaction 
the overall symmetry of the unperturbed Skyrmions gets broken generically, 
even for the isolated Skyrmions and  these coupled set of
equations, unlike in  the pure ferromagnetic model, can not be solved 
analytically. Nevertheless, the linearity of these equations allows  numerical
solutions for the fields $\beta_1 , \gamma_1$ (see Supplementary material
for Mathematica 10 code). Adding now these  perturbing 
solutions for the DM interaction to the corresponding analytic solutions $\beta_0 , \gamma_0$
found above for the ferromagnetic model, which also stabilizes the
solutions,
 we  finally obtain the Skyrmions for the helimagnetic model in the form (\re
{Pbetgam}) as shown in Fig.  2, for both isolated and crystal solutions.

\begin{figure}[h!]
 \includegraphics
[width=6cm,height=4 cm]
{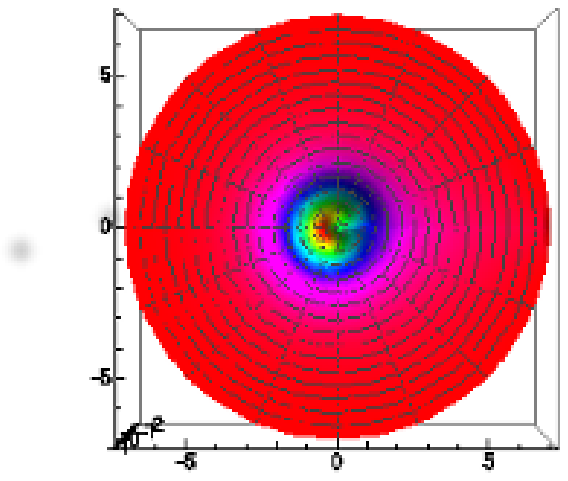}
\quad \quad \quad
 \includegraphics
[width=6.cm,height=4. cm]
{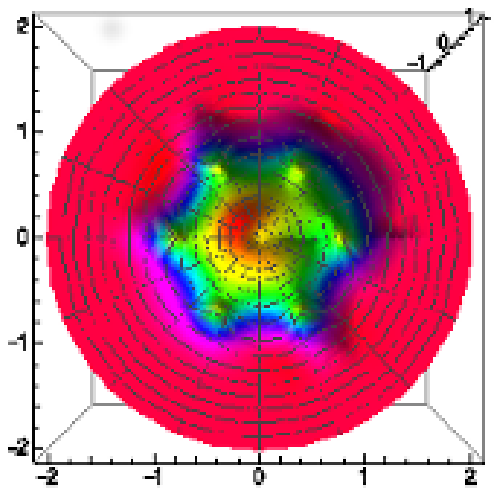}
\caption{ Magnetization component $M^3=\cos \beta $ in helimagnetic  model
 (\ref{FM}-\re{DM}) for a) individual Skyrmion with $Q=-1 $ and b) hexagonal  Skyrmion
crystal with $Q=-7 $. These Skyrmion solutions are   perturbations of
ferromagnetic Skyrmions obtained in Fig. 1  for weak DM coupling, with small parameter
$\epsilon=0.2$.   }
\end{figure}

It is evident by comparing the Skyrmion  solutions 
for  the Helimagnetic
model presented in  Fig. 2 with those for the pure ferromagnetic model shown in Fig. 1, that 
the inclusion of the symmetry breaking DM
interaction (\ref{DM}) deforms the 
 solutions and  introduces   further
 anisotropy and noncircular symmetry to the
 individual  Skyrme molecules  as well as to  the  Skyrme crystals appearing
 in the helimagnet model. This result generalizes therefore  the known  circular
 symmetric Skyrmions  and Skyrmion crystals obtained  in the helimagnetic model
 to the asymmetric case. 
\section {Concluding remarks}
  Appearance of   Skyrme crystals   in the   hexagonal form 
has been  observed experimentally in the  magnetization 
pattern  in  helimagnetic materials. However a full   
theoretical description for the formation of   crystalline structures  
   as a general topological Skyrmion
solution of  the governing equations minimizing the model Hamiltonian, which is
difficult to achieve due to high asymmetry and nonlinearity inherent to
the system is not yet available in the literature.
 Assuming weaker DM interaction relative to the ferromagnetic
exchange coupling, which is also a requirement  for the 
conventional continuum approximation, we present here such  solutions
 for  Skyrme crystals with higher topological charge and for
individual Skyrmions with unit charge in the helimagnetic model
. Our
approach of perturbing and stabilizing the  Skyrmions   of the ferromagnetic model
 with
the symmetry breaking DM interaction, bypasses the major difficulty of
nonlinearity and solves  the energy minimizing equations 
 without assuming any separation of
variables or circular cell approximation
, for the first
time (see  the Supplementary material).
Note that, though  in  recent   experiments the evidence of 
Skyrmion and Skyrme crystals was found, the details about the 
circular symmetry or isotropy of  Skyrmions in their  individual state or when they
appear as interacting molecules in the Skyrme crystal are not much clear
(see color-code of  experimental
images in \cite{LorentzExp})
 Therefore,  the theoretical possibility of  asymmetric Skyrme
crystals found here,  can not be ruled out
without  verifying 
in precision experiments, where deformation of symmetries and isotropy in the magnetic
pattern  could  be detected 
in finer details.
  To tune  with the validity region of
the present
 solutions obtained for low  values of $\epsilon $,  care  should be taken  
in choosing the helimagnetic materials, so that their spin-orbital
   DM interaction is   weaker in  comparison  with  the ferromagnetic
spin exchange. Though the helimagnetic crystals 
MnSi and Fe$_{1-x}$Co$_x$Si  share similar magnetic
features, they differ significantly in their electronic structures, which
is likely to make the later system  more suitable for investigating the   nature
of Skyrmion lattices over a large temperature range, with the possibility of
revealing more refined structures. 
The recent Lorentz TEM experiment \c{LorentzExp}
might well be revisited and rescrutinized for detecting the possible symmetry breaking
in the observed magnetic pattern in Skyrme crystals, 
in the light of the present result, at least in the excited states and  in the
higher temperature range.

\end{document}